\documentclass[
    ,final            
  ,numberedheadings 
  ]
  {aipproc}

\layoutstyle{8x11single}

\begin{document}

\title{Correlative Spectral Analysis of Gamma-Ray Bursts using Swift-BAT and GLAST-GBM\footnote{Adapted from a contribution to the Proceedings of the 2008 Nanjing GRB Conference. Edited by Y. F. Huang, Z. G. Dai and B. Zhang.}}


\author{Michael Stamatikos\footnote{Correspondence to Michael.Stamatikos-1@nasa.gov.}      $^{,}$}{
  address={Astroparticle Physics Laboratory, Code 661, NASA/Goddard Space Flight Center, Greenbelt, MD 20771 USA}
}

\author{Taka Sakamoto}{
  address={Astroparticle Physics Laboratory, Code 661, NASA/Goddard Space Flight Center, Greenbelt, MD 20771 USA},altaddress={CRESST, University of Maryland, Baltimore County, Baltimore, MD 21250 USA}
}

\author{David L. Band}{
  address={Astroparticle Physics Laboratory, Code 661, NASA/Goddard Space Flight Center, Greenbelt, MD 20771 USA},altaddress={CRESST, University of Maryland, Baltimore County, Baltimore, MD 21250 USA}
}

\begin{abstract}
We discuss the preliminary results of spectral analysis simulations involving anticipated correlated multi-wavelength observations of gamma-ray bursts (GRBs) using Swift's Burst Alert Telescope (BAT) and the Gamma-Ray Large Area Space Telescope's (GLAST) Burst Monitor (GLAST-GBM), resulting in joint spectral fits, including characteristic photon energy $\left(E_{peak}\right)$ values, for a conservative annual estimate of $\sim30$ GRBs. The addition of BAT's spectral response will (i) complement in-orbit calibration efforts of GBM's detector response matrices, (ii) augment GLAST's low energy sensitivity by increasing the $\sim20-100$ keV effective area, (iii) facilitate ground-based follow-up efforts of GLAST GRBs by increasing GBM's source localization precision, and (iv) help identify a subset of non-triggered GRBs discovered via off-line GBM data analysis. Such multi-wavelength correlative analyses, which have been demonstrated by successful joint-spectral fits of Swift-BAT GRBs with other higher energy detectors such as Konus-WIND and Suzaku-WAM, would enable the study of broad-band spectral and temporal evolution of prompt GRB emission over three energy decades, thus potentially increasing the science return without placing additional demands upon mission resources throughout their contemporaneous orbital tenure over the next decade.
\end{abstract}

\maketitle


\section{Motivation: Joint BAT/GBM GRB Observations}

The Swift MIDEX explorer mission \cite{Gehrels:2007} (anticipated to operate until at least $\sim2014$), comprised of the wide-field ($\sim1.4$ sr, half-coded) hard X-ray (15-150 keV) Burst Alert Telescope (BAT) \cite{Barthelmy:2005}, and the narrow-field (0.2-10 keV) X-Ray (XRT) and (170-600 nm) Ultraviolet-Optical (UVOT) Telescopes, has revolutionized our understanding of GRBs. The intrinsic multi-wavelength instrumentation, coupled with a rapid ($<\sim$100 seconds) autonomous slew capability, has ushered in an unprecedented era of source localization precision $\left(<\sim1^{\prime}-4^{\prime}\right)$ that is disseminated in real-time ($\sim10$ seconds) via the GRB Coordinate Network (GCN), which enables broad-band international observational campaigns. Swift's unique dynamic response, in conjunction with correlative ground-based follow-up efforts, has resulted in $\sim70$ observed spectroscopic redshifts, while its X-ray sensitivity has revealed evidence of extended soft emission in some short GRBs.

The Gamma-Ray Large Area Space Telescope, which is now known as the Fermi Gamma-Ray Space Telescope (\emph{Fermi}), mission was launched on June 11, 2008 and has an anticipated orbital lifetime spanning until at least $\sim2018$. A goal of GLAST, which is comprised of the (20 MeV to $>$300 GeV) Large Area Telescope (LAT) and the (8 keV - 30 MeV) GLAST Burst Monitor (GBM), which is now known as the GRB Burst Monitor (GBM), is to study transient gamma-ray sources, while a direct GBM experimental objective is to identify and study GRBs. Common scientific interest between GLAST and Swift, in the context of prompt GRB emission, provides strong motivation for a cross-calibration via correlative observations of GRBs, resulting in joint spectral energy fits, thus enabling the analysis of multi-wavelength spectral and temporal evolution. Cross-calibration with BAT would also complement in-orbit calibration efforts of GBM's detector response matrices and would be especially critical during GLAST's first year, since it would facilitate a smooth transition from performance diagnostics to science operations.

\begin{figure}
\includegraphics[height=.54\textheight]{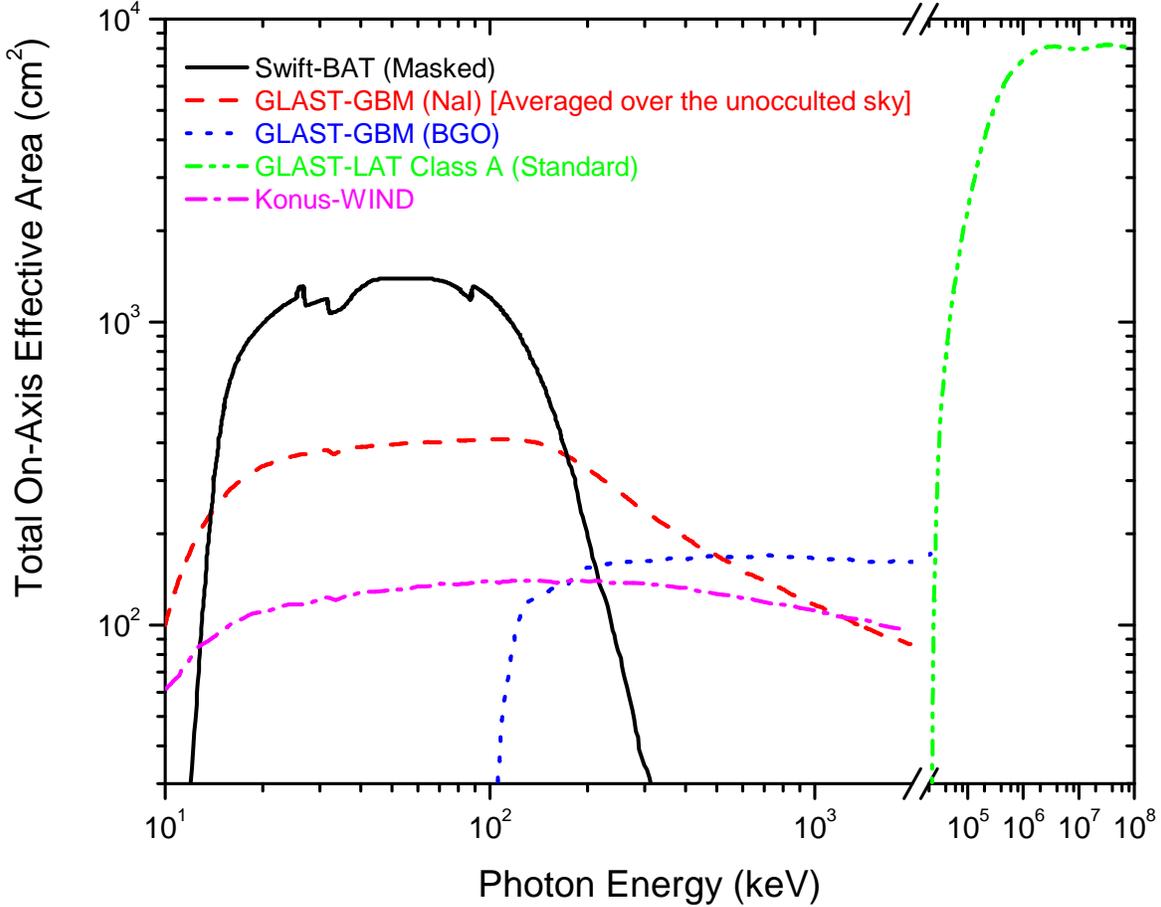}
\caption{Effective areas for Swift-BAT, GBM-NaI, GBM-BGO, GLAST-LAT and Konus-WIND. \emph{Note the break in the x-axis.}}
\label{Overlapped_Effective_Areas}
\end{figure}

The GBM, consisting of 12 NaI (8-1000 keV) and 2 BGO (0.15-30 MeV) detectors, will monitor $\sim8$ steradians of the sky, and, in concert with LAT, enables GLAST to continuously span 7 energy decades, but not at the same sensitivity. As illustrated in Figure~\ref{Overlapped_Effective_Areas}, GLAST's effective area drops by over $\sim1.5$ orders of magnitude from LAT (GeV) to GBM-NaI (keV) energies, while the (masked) BAT ($\sim$20-100 keV) low energy effective area surpasses GBM-NaI's by over a factor of $\sim3$. Furthermore, although Swift has detected $>\sim300$ GRBs, the majority of E$_{peak}$ (characteristic photon energy) values lie beyond BAT's canonical energy range, since $\overline{E}_{peak}\sim250$ keV. Thus, correlated Swift-BAT/GLAST-GBM GRB observations would simultaneously augment GLAST's low energy response while increasing the number of $E_{peak}$ for BAT GRBs observed by GBM. Additionally, since Swift's high fidelity localization precision surpasses GBM's by over $\sim2-3$ orders of magnitude, we expect that $\sim30\%$ of bursts in the joint BAT-GBM analysis would be accompanied by panchromatic ground-based follow-up observations resulting in observed spectroscopic redshifts and host galaxy identifications.

\section{Methodology: GRB Overlap \& Simulation Studies}

The annual GRB trigger rates for BAT and GBM are $\sim100\pm10$ and $\sim200\pm20$, respectively. Assuming isotropic GRB spatial distribution with uncorrelated FOVs over homogenous sky coverage, we expect the number of BAT GRBs within GBM's FOV $\left(n_{GBM}^{BAT}\right)$ and the number of GBM GRBs within BAT's FOV $\left(n_{BAT}^{GBM}\right)$ to be $\sim64\pm10$ and $\sim21\pm16$, respectively. Due to BAT's superior sensitivity, we assume that all $n_{BAT}^{GBM}$ would trigger Swift. Hence, $n_{BAT}^{GBM}$ serves as a conservative lower limit for the annual number of GRBs anticipated for BAT-GBM cross-calibration $\left(n_{GRB}^{XCal}\right)$.

Based upon the implicit degeneracy, we expect the number of $n_{GBM}^{BAT}$ failing to trigger GBM $\left(n_{GBM}^{BAT\prime}\right)$ to be $\sim43\pm19$, i.e. we expect only $\sim33\%\pm26\%$ of $n_{GBM}^{BAT}$ will trigger GBM-NaI on-board. It is planned that GBM count rate data will be continuously down-linked in 8 energy channels at 256 ms temporal resolution and in 128 energy channels at 4 s temporal resolution, making it possible to detect additional non-triggered GRBs off-line. We estimate that an additional $n_{GBM}^{GND\prime}\sim80\pm8$ GRBs will be found on the ground (GND). Hence, GBM spectra may still be available in the absence of an on-board GLAST trigger. However, $n_{GBM}^{GND\prime}$ would suffer from reduced temporal and spectral resolution, with backgrounds that will be difficult to quantify for low peak flux, long duration GRBs. Selection effects, such as detector composition and long accumulation timescales, bias BAT towards long, soft GRBs with lower $E_{peak}$ \cite{Band:2008d}. Consequently, BAT GRBs comprise a separate statistical class, as is demonstrated by their fluence and redshift distributions \cite{Band:2006}, from classical Burst and Transient Source Experiment
BATSE GRBs, which have been used to estimate the GBM on-board detection rate due to the similarities between GBM and BATSE. Hence, BAT may facilitate off-line GRB efforts since a subset of $n_{GBM}^{GND\prime}$ may be populated by $n_{GBM}^{BAT\prime}$.

Two sets $\left(\texttt{X=A,B}\right)$ of simulations were performed using 175 BAT GRBs in order to estimate the detection $\left(\epsilon_{t_{\texttt{X}}}^{NaI}\equiv n_{\sigma_{t}\geq5}^{NaI}/n_{GRB}^{Set \texttt{ X}}\right)$ and spectral $\left(\epsilon_{s_{\texttt{X}}}^{NaI,BGO}\equiv n_{\sigma_{s}\geq15}^{NaI,BGO}/n_{GRB}^{Set \texttt{ X}}\right)$ efficiencies of $n_{GBM}^{BAT\prime}$ in GBM and $n_{BAT}^{GBM}$ in BAT. Set $\texttt{A}$ consisted of $n_{GRB}^{Set \texttt{ A}}=165$ Swift GRBs whose spectral parameters were estimated using a combination of the best BAT fits and an empirical mapping between the fitted low energy spectral index and $E_{peak}$. Set $\texttt{B}$ consisted of $n_{GRB}^{Set \texttt{ B}}=27$ intense Swift GRBs whose energy spectral fit parameters were taken from Band function fits \cite{Band:1993} to correlated Konus-WIND data, and functions as a surrogate for $n_{BAT}^{GBM}\sim21\pm16$.

\begin{figure}
\includegraphics[height=.54\textheight]{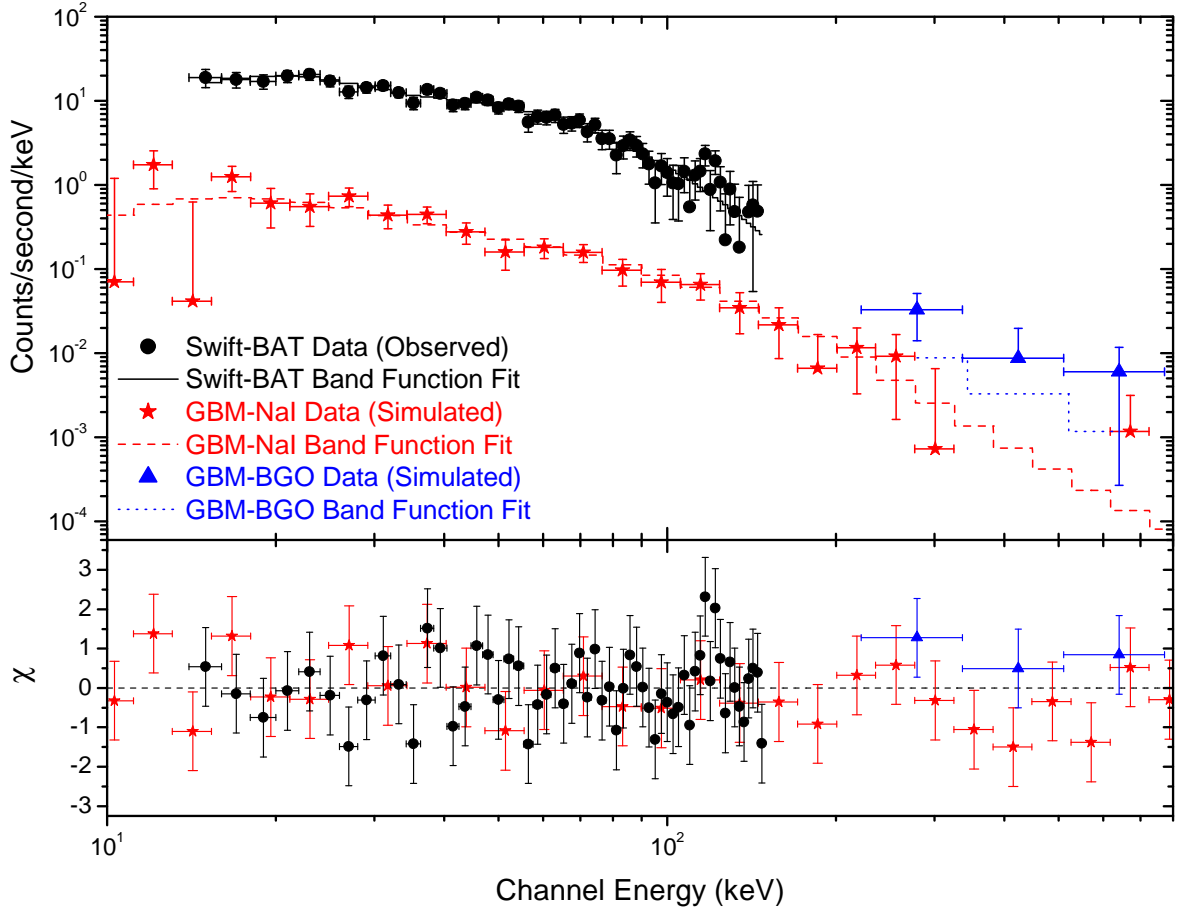}
\caption{Simulated joint spectral Band function fit for GRB 060204B using Swift-BAT, GLAST-GBM-NaI and GLAST-GBM-BGO data, which resulted in an E$_{peak}=100\pm37$ keV.}
\label{GRB060204B_BAT_GBM_Joint_Spectral_Fit}
\end{figure}

\section{Results: Extrapolated Joint BAT/GBM Fits}

$\texttt{GLASTspec}$\footnote{An $\texttt{XSPEC}$ interface $\texttt{(http://heasarc.gsfc.nasa.gov/webspec/GLASTspec.html)}$.} was used to extrapolate the spectra of sets $\texttt{A}$ and $\texttt{B}$ to the GBM energy range. Using the detection criteria of $\sigma_{t}^{NaI}\geq5$ in the $\Delta E=50-300$ keV band, on timescales of up to $\Delta t=4$ seconds, in at least 2 NaI detectors, and $\sigma_{s}^{NaI,BGO}\geq15$ ($\Delta E_{NaI}=8-1040$ keV, $\Delta E_{BGO}=142.5-30,892$ keV) as a proxy for good spectral quality \cite{Ford:1995}, it was determined that $\epsilon_{t_{\texttt{A}}}^{NaI}\sim32\%$, in agreement with expectations, while $\sim19\%$ of the GRBs in set $\texttt{A}$ satisfied the inequalities: $\sigma_{t}^{NaI}<5$ and $\sigma_{s}^{NaI}\geq15$. A reduction of the spectral quality criterion to $\sigma_{s}^{NaI}\geq10$ increased the number of non-triggered NaI spectra to $\sim34\%$. We use this range to estimate the spectral efficiency of $n_{GBM}^{BAT\prime}$ $\left(\epsilon_{s}^{NaI\prime}\equiv n_{\sigma_{s}\geq10}^{NaI\prime}/n_{GBM}^{BAT\prime}\sim0.265\pm0.075\right)$, since bursts in this subset would yield adequate NaI spectra off-line, while failing to trigger GBM on-board. The simulation results indicate that for $10<\sim\sigma_{s}^{NaI}<\sim15$, E$_{peak}$ may still be determined via joint fits, even when unilateral BAT or GBM spectral analysis was inadequate, thus underscoring the necessity of joint fits, as illustrated in Figure~\ref{GRB060204B_BAT_GBM_Joint_Spectral_Fit} for a simulated BAT/GBM joint spectral fit of GRB 060204B, with $\sigma_{t}^{NaI}\sim2$ and $\sigma_{s}^{NaI}\sim13$, which resulted in fit parameters of $\alpha=-0.89_{-0.26}^{+083}$, $\beta=-2.39_{-7.61}^{+0.62}$ and  $E_{peak}=100\pm37$ keV, for a $\chi/dof=75/101$.

Meanwhile, we found that all GRBs in set $\texttt{B}$ would have triggered and produced good quality spectra in NaI, with $\sim70\%$ also registering good quality spectra in BGO. This illustrates that our initial assumptions regarding the high triggering and spectral efficiency of $n_{BAT}^{GBM}$ are reasonable, since the effective area of Konus-WIND is a factor of $\sim3$ smaller than that of NaI in the BAT energy range (see Figure~\ref{Overlapped_Effective_Areas}). The application of $\epsilon_{s}^{NaI\prime}$ onto $n_{GBM}^{BAT\prime}$ gives the estimated number of non-triggered BAT GRBs in GBM's FOV that would render good quality spectra in NaI as $n_{GBM}^{BAT\prime\prime}\sim11\pm6$, which means that $\sim(14\pm8)\%$ of $n_{GBM}^{GND\prime}$ could be identified by BAT. Consequently, we estimate that $n_{GRB}^{XCal}\approx n_{BAT}^{GBM}+n_{GBM}^{BAT\prime\prime}\sim32\pm17$, which is similar to the overlap observed with other high energy detectors. We note that these conservative estimates, tantamount to a few BAT-GBM GRB cross-calibrations per month, can be improved by correlating BAT-GBM FOVs via maximizing the overlap of their respective sky pointing directions.

\section{Discussion \& Future Outlook}

Current efforts include inter-calibrating BAT and GBM, in collaboration with the GBM team, via joint spectral fits of mutually observed GRBs. Previous studies of joint-fit GRB data sets \cite{Krimm:2006c} enhanced our understanding of burst parameter classifications, explored GRB emission geometry, and tested the viability of various redshift estimation methods \cite{Stamatikos:2008c}. In addition, a more accurate normalization between GLAST prompt emission and Swift afterglow spectra will facilitate the determination of GRB energy budgets. Overall, broad-band correlative studies will enable the investigation of spectral and temporal evolution \cite{Stamatikos:2008e} over unprecedented decades of energy, shedding light on the connection between electromagnetic pulse asymmetry, width and spectral softening \cite{Norris:1996}, while facilitating multi-messenger searches such as those for correlated leptonic emission \cite{Stamatikos:2005,Stamatikos:2006b}. Hence, correlated Swift-BAT/GLAST-GBM observations enhance their science return and benefit the broader astronomical community, without additional demands upon mutual mission resources, thus underscoring an aspect of Swift's operational and scientific relevancy in the impending era of GLAST.



\begin{theacknowledgments}
The authors are grateful to the members of the GBM team for very fruitful discussions in regards to this analysis. M. Stamatikos is supported by an NPP Fellowship at NASA-GSFC administered by ORAU.
\end{theacknowledgments}



\bibliographystyle{aipproc}   

\bibliography{Stamatikos_GRB_Nanjing_astroph2}

\hyphenation{Post-Script Sprin-ger}
\begin{thebibliography}{12}
\expandafter\ifx\csname natexlab\endcsname\relax\def\natexlab#1{#1}\fi
\providecommand{\enquote}[1]{``#1''}
\expandafter\ifx\csname url\endcsname\relax
  \def\url#1{\texttt{#1}}\fi
\expandafter\ifx\csname urlprefix\endcsname\relax\def\urlprefix{URL }\fi
\providecommand{\eprint}[2][]{\url{#2}}

\bibitem[{Gehrels} et~al.(2007)]{Gehrels:2007}
N.~{Gehrels}, J.~K. {Cannizzo}, and J.~P. {Norris}, \emph{New Journal of
  Physics} \textbf{9}, 37 (2007).

\bibitem[Barthelmy et~al.(2005)]{Barthelmy:2005}
S.~D. Barthelmy, et~al., \emph{Space Science Reviews} \textbf{120}, 143--164
  (2005).

\bibitem[Band(2008)]{Band:2008d}
D.~L. Band, \emph{AIP Conference Series} \textbf{1000}, 121--124 (2008).

\bibitem[Band(2006)]{Band:2006}
D.~L. Band, \emph{The Astrophysical Journal} \textbf{644}, 378--384 (2006),
  \eprint{astro-ph/0602267}.

\bibitem[Band et~al.(1993)]{Band:1993}
D.~L. Band, et~al., \emph{The Astrophysical Journal} \textbf{413}, 281--292
  (1993).

\bibitem[Ford et~al.(1995)]{Ford:1995}
L.~A. Ford, et~al., \emph{The Astrophysical Journal} \textbf{439}, 307--321
  (1995), \eprint{astro-ph/9407090}.

\bibitem[{Krimm} et~al.(2006)]{Krimm:2006c}
H.~A. {Krimm}, et~al., \emph{AIP Conference Series} \textbf{836}, 145--148
  (2006).

\bibitem[Stamatikos et~al.(2008)]{Stamatikos:2008c}
M.~Stamatikos, et~al., \emph{AIP Conference Series} \textbf{1000}, 137--141
  (2008).

\bibitem[{Stamatikos} et~al.(2008)]{Stamatikos:2008e}
M.~{Stamatikos}, T.~N. {Ukwatta}, T.~{Sakamoto}, and K.~S. {Dhuga}, \emph{ArXiv
  e-prints} \textbf{809} (2008), \eprint{0809.2132}.

\bibitem[Norris et~al.(1996)]{Norris:1996}
J.~P. Norris, et~al., \emph{The Astrophysical Journal} \textbf{459}, 393--412
  (1996).

\bibitem[Stamatikos et~al.(2005)]{Stamatikos:2005}
M.~Stamatikos, J.~Kurtzweil, and M.~J. Clarke, \emph{Proceedings of the
  29$^{th}$ ICRC} \textbf{4}, 471--474 (2005), \eprint{astro-ph/0510336}.

\bibitem[Stamatikos and Band(2006)]{Stamatikos:2006b}
M.~Stamatikos, and D.~Band, \emph{AIP Conference Series} \textbf{836}, 599--604
  (2006), \eprint{astro-ph/0602481}.

\end{thebibliography}

\IfFileExists{\jobname.bbl}{}
 {\typeout{}
  \typeout{******************************************}
  \typeout{** Please run "bibtex \jobname" to optain}
  \typeout{** the bibliography and then re-run LaTeX}
  \typeout{** twice to fix the references!}
  \typeout{******************************************}
  \typeout{}
 }

\end{document}